\documentclass[10pt]{article}

\usepackage{amsmath}
\usepackage{amssymb}

\usepackage{graphicx}

\usepackage{cite}

\usepackage{color} 

\usepackage{todonotes}

\usepackage[labelfont=bf,labelsep=period,justification=raggedright]{caption}

\makeatletter
\renewcommand{\@biblabel}[1]{\quad#1.}
\makeatother

\date{}

\DeclareMathOperator*{\argmin}{arg\,min}

\begin{document}

\begin{flushleft}
{\Large
\textbf{A Generic Approach to Solving Jump Diffusion Equations with Applications to Neural Populations
}
}
\\
Marc de Kamps$^{1}$
\\
\bf{1}  School of Computing, University of Leeds, Leeds, West Yorkshire, United Kingdom
\\
$\ast$ E-mail: Corresponding M.deKamps@leeds.ac.uk
\end{flushleft}

\section*{Abstract}

Diffusion processes have been applied with great success to model the dynamics of
large populations throughout science, in particular biology. One advantage is that they bridge two different scales: the microscopic and the macroscopic one. Diffusion is a mathematical idealisation, however: it assumes vanishingly small state changes at the microscopic level. In real biological systems this is often not the case. The differential Chapman-Kolmogorov equation is more appropriate to model population dynamics that is not well described by drift and diffusion alone.

Here, the method of characteristics is used to transform deterministic dynamics away and find a coordinate frame where this equation reduces to a Master equation. There is no longer a drift term, and solution methods there are insensitive to density gradients, making the method suitable for jump processes with arbitrary jump sizes.  Moreover, its solution is universal: it no longer depends explicitly on the deterministic system. 

We demonstrate the technique on simple models of neuronal populations. Surprisingly,
it is suitable for fast neural dynamics, even though in the new coordinate
frame state space may expand rapidly towards infinity. We demonstrate universality: the method is applicable to any one dimensional neural model and show this on populations of leaky-
and quadratic-integrate-and-fire neurons. In the diffusive limit, excellent approximations of Fokker-Planck equations are achieved. Nothing in the approach is particular to neuroscience, 
and the neural models are simple enough to serve as an example of dynamical systems that
demonstrates the method.

\section*{Author Summary}
Biological systems stand out by their complexity and multi-scale interactions. Modeling techniques
that can explain the behavior of systems - the macroscopic level - in terms of their individual components 
- the microscopic level - are very valuable. Diffusion models do that, so, unsurprisingly, 
they pervade biology. One key assumption underlying the
validity of the diffusion approach is that state changes of individuals are minuscule. In real biological
systems this can often not  be justified. The method presented here dispenses with the
assumption and broadens the application range of population modelling considerably. Moreover,
the method reduces the mathematically difficult problem of solving a partial differential equation 
to a simpler one: understanding the noise process that gave rise to it. 

We demonstrate the approach
on neuronal populations. We envisage that the study of networks of such populations will elucidate brain function,
as large-scale networks can be simulated with unprecedented realistic neural dynamics. This may
help explain imaging signals, which after all are the observable collective signal of large groups of neurons,
or it may help researchers to develop their ideas on neural coding. We also expect
that the technique will find broad application outside of neuroscience and biology, in particular in finance.

\section*{Introduction}
Diffusion models are of crucial importance to biological modeling; literally, text books have been filled on the 
subject, e.g. \cite{Ricciardi1977,murray2002}.  The application domain is enormous: fluid dynamics, chemotaxis
\cite{alt1980}, animal population movement \cite{skellam1951,skellam1973}, neural populations e.g.
\cite{stein1965,   knight1972, omurtag2000, brunel2001, haskell2001, casti2002, fourcauld2003, mattia2004,  apfaltrer2006, richardson2007, helias2011, dumont2012}  are just a few examples (see \cite{codling2008} for a recent review
and further references). Diffusion models allow an understanding of group 
behaviour in terms of that of its constituents, i.e. they link macroscopic and microscopic behaviour. In 
general diffusion models emerge as follows: microscopic deterministic laws describing individual behavior in the 
absence of interactions are determined. Interactions among individuals and between individuals and
the outside world are described by a stochastic process ('noise'). Diffusion arises if one assumes that changes 
in the microscopic state as a consequence of an interaction are vanishingly small. Sometimes, this assumption 
is well justified: for example, in Brownian motion these interactions are due to the collisions of individual 
molecules, interactions that are indeed minuscule compared to the macroscopic state. Often, this assumption is 
not  appropriate, but made in order to use the familiar mathematical machinery of diffusion equations. 
For example, in  computational neuroscience the diffusion approximation 
hinges on the assumption that synaptic efficacies - a measure for how strongly neurons influence each other - 
are small. This is simply not the case in many brain areas, e.g. \cite{sirovich2007}, and leads to sizable
corrections to diffusion results \cite{helias2010}. Nevertheless, the diffusion approximation predominates
in computational neuroscience. Given the widespread distribution of diffusion models in 
biology, it stands to reason that this situation is common.

Experience from computational neuroscience \cite{helias2010} and finance - a field that has put considerable 
effort into understanding jump diffusion equations e.g. \cite{merton1976,kou2002,cont2005} - has shown that 
results valid beyond the diffusion limit are very hard to obtain, limited to a specific study and often require a 
perturbation approach \cite{helias2010}. Due to their nature, jump diffusion processes may lead to very jagged, 
locally discontinuous probability density profiles \cite{gardiner1997}. For this reason numerical solution schemes 
that are well established for diffusion processes may be unsuitable for finite size jumps. 
Some solution schemes are applicable for finite jumps if it can be assumed that the underlying density profile is 
smooth. Brennan and Schwartz \cite{brennan1978}  demonstrated that an implicit scheme for solving the Black-Scholes
 equation is equivalent to a generalized jump process.  The assumption must be justified, however. In contrast,
the  method presented here makes
no assumptions about the structure of the underlying structure of the density profile at all.
Below, we will present an example that can be easily modeled by the method presented here, but where we are not 
aware of more conventional discretization schemes that would apply.

Standard text books such as \cite{gardiner1997,vankampen1992} demonstrate how deterministic laws of motion for 
individuals and a Master equation describing the noise process under consideration can be combined. 
This results in the so-called  differential Chapman-Kolmogorov equation, a partial integro-differential equation 
in probability density (it is derived in the context of neuronal populations in {\bf ``Methods: Derivation of the population density equation''}). When the assumption is made that the transitions due to the stochastic processes
induce  very small state transitions, this equation reduces to a partial differential equation of the 
Fokker-Planck type \cite{gardiner1997,vankampen1992}. Diffusion models ultimately derive from this equation. 
Clearly, a solution method for the Chapman-Kolmogorov equation is highly valuable: 
it would still allow the study of diffusion processes, but also processes with large jumps. Below,
we will give an example of the gradual breakdown of diffusion. This opens up an important application area:
it will be possible to re-examine results already obtained in the diffusion limit and to investigate the 
consequences of finite jump sizes.

Many expositions immediately proceed to the diffusion regime. Here we follow
a radically different approach: we use the method of characteristics to define a coordinate frame that co-moves 
with the flow of the deterministic
system. In this frame the differential Chapman-Kolmogorov equation
reduces to the Master equation of the stochastic process under consideration. This approach neatly
sidesteps the need for solving  a partial (integro-) differential equation as the Master equation is a set of 
ordinary differential equations, and can be solved by relatively simple numerical methods. 

This simplicity comes at a price, however: the probability density  is represented in an  interval 
that is moving with respect to the new coordinate frame. How this interval behaves becomes dependent 
on the topology of the flow of the deterministic system. We will illustrate the problem on two neuron 
models that represent two ends of the dynamic spectrum. Leaky-integrate-and-fire
neurons have slow underlying deterministic dynamics. Quadratic-integrate-and-fire neurons model the rapid 
onset of a spike. The dynamics is so fast that the original interval representing the 
probability density profile moves to infinity in finite time in the
new coordinate frame. 

We find that these neural models are a demonstration of the broad applicability of the technique. Not only are
they simple examples of dynamical systems that can easily be understood without a background in computational 
neuroscience. They also provide boundary conditions that any good solution method must be able to 
handle: absorbing boundaries. In general neurons spike and then return to pre-spike conditions. 
In one dimensional neural
models this is simulated by resetting the neuronal state as soon as an absorbing boundary (the 'threshold')
is reached. The reintroduction of a neuron once it has hit the threshold (a 'reset') is particular to 
neural systems. It is a marked difference from financial derivatives - as options become worthless after
their expiration date - and one reason why results from finance do not carry over to neuroscience despite
similarities.

The method is manifestly stable, and the jump size is immaterial. The method can be considered a generalization
of the diffusion approach: for a stochastic process characterized by small jumps diffusion  results are recovered,
as one would expect. However, it  is not always practical to study diffusion by a small finite jump process as the
convergence to diffusion results is not always uniform and a direct application of the Fokker-Planck equation
may then give better results. We believe the main application of the method is for processes with truly finite 
jumps, or to study deviations from the diffusion limit.

\section*{Results}
We obtained the following results.
\begin{enumerate}
  \item The method of characteristics was employed to define a new coordinate frame that moves along the flow of 
  deterministic dynamical system. In the new coordinate frame the differential Chapman-Kolmogorov equations transforms
  to the time dependent Master equation of the noise process. This result was used earlier in the limited
  setting of leaky-integrate-and-fire neurons \cite{dekamps2003,dekamps2006}. It is restated here for convenience.

  \item The characteristics themselves were used to define a representation of the probability density. This
    is a key new insight as it allows an accurate probability representation even when the deterministic process 
    is fast. This is explained in some detail below.

  \item The density representation is dependent on the topology of the deterministic flow. Although the number
    of topologies is typically limited, potentially this could require a novel implementation of the algorithm.
    Somewhat surprisingly, we found that the simplest implementation of the density is for a deterministic
    process that corresponds to periodically spiking neurons. Importantly, we found that under broad conditions
    it is possible to 'switch topology' by a process we dubbed 'current compensation'. It is possible
    to modify the topology of the deterministic neural dynamics by adding a DC current to the neurons. 
     We compensate
    by renormalizing the noise input such that its mean is lower by an amount equal to the DC current.
    Therefore, often a single topology suffices.

  \item 
    The algorithm that results from the application of the technique is independent of the neural model. 
     Exploring different models does not require recoding of the algorithm, but merely the recalculation of a grid
    representation. The algorithm can be considered to be a universal population density solver for one
    dimensional neural models whenever current compensation is appropriate.
    We delivered an implementation of the algorithm in C++ that is publicly available (http://miind.sf.net).
    We used this implementation in the examples that illustrate the use of this algorithm.
    
\end{enumerate}

\subsection*{Transforming away neural dynamics}

Assume that a neuron  is characterised by a vector $\vec{v} \in M$ summarising its state, where $M$ is a open 
subset of $\mathbb{R}^n$. Further, assume that a smooth vector field $\vec{F}(\vec{v})$ exists everywhere on $M$ 
and that a density function $\rho(\vec{v},t)$ is defined for every $\vec{v} \in M$. Now consider a large 
population of neurons. 
$\rho(\vec{v} )d \vec{v}$ is the probability for a neuron to have its state vector in $d \vec{v}$. 
It is also assumed that the population is homogeneous in the sense that interactions between neurons and the 
outside world can be accounted for by the same stochastic process for all neurons (although individual neurons 
each see different realisations of this process). In the absence of noise, neurons
follow trajectories through state space determined by the flow of vector field $\vec{F}(\vec{v})$:
\begin{equation}
  \tau \frac{d \vec{v} }{dt} = \vec{F}(\vec{v}),
  \label{eq-dyn}
\end{equation}  
where $\tau$ is the neuron's time constant. We must allow for the possibility of $v$ being driven across
$\partial M$, the edge of $M$, at which time it must be reset, immediately or after a refractive period at 
$v=V_{reset}$.  
   
As explained in {\bf ``Methods: Derivation of the Population Density Equation''}, conventional balance arguments 
lead to an advection equation for the density. When a noise process is also taken into consideration, 
this equation becomes:
\begin{equation}
  \frac{ \partial \rho }{ \partial t } + \frac{\partial}{ \partial \vec{v}} \cdot ( \frac{\vec{F} \rho}{\tau} ) = 
\int_{M} d \vec{w} \left\{ W(\vec{v} \mid \vec{w})\rho(\vec{w}) - W(\vec{w} \mid \vec{v}  )\rho(\vec{v}) \right\} 
\label{eq-balance}
\end{equation} 
$W(\vec{w} \mid \vec{v} )$ is the probability per unit time for a noise event that will cause a state transition 
from  $d \vec{v} $ to $d \vec{w}$. 
This equation is known as the \emph{differential Chapman-Kolmogorov equation} \cite{gardiner1997}.

\begin{figure}[!ht]
\begin{center}
  \includegraphics[width=0.9\textwidth]{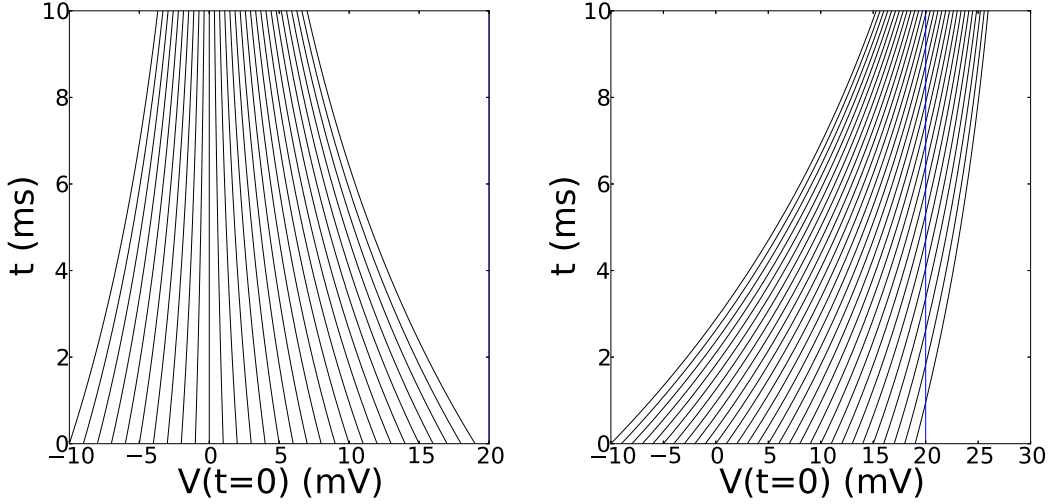}
  \caption{{\bf Solutions to Eq. \ref{eq-dyn} for LIF neurons ($V_{th} = 20$ mV, $\tau = 10$ ms) for $I = 0$ mV (left) or $I = 30$ mV (right).} }
  \label{fig-lifchar}
\end{center}
\end{figure}

Under the usual assumptions that guarantee the existence and uniqueness of solutions of Eq. \ref{eq-dyn} on $M$, 
one can find integral curves $\vec{v}^{\prime}(t,\vec{v_0}(t_0))$. 
If $\vec{v_0}$ and $t_0$ are suitably chosen on $\partial M$,  these curves cover the entire manifold $M$
and every point of $M$ is uniquely defined by a coordinate pair $(\vec{v}^{\prime},t)$. 

These curves are the characteristics of Eq. \ref{eq-balance}. Applying the method of characteristics one finds that
the total time derivative 
of the density in the $(\vec{v}^{\prime},t)$ system is given by:

\begin{equation}
  \frac{d \rho^{\prime}}{dt} =\int_{M^{\prime}} d \vec{w^\prime} \left\{ W(\vec{v^\prime} \mid \vec{w^\prime})\rho^{\prime}(\vec{w^\prime}) - 
W(\vec{w^\prime} \mid \vec{v^\prime}  )\rho^{\prime}(\vec{v^\prime}) \right\} 
\label{eq-master}
\end{equation}
where $\rho^{\prime}(\vec{v}^{\prime},t) = e^{\int^t \frac{\partial F(\vec{v}^{\prime})}{\tau \partial \vec{v}^\prime}dt^\prime} \rho(\vec{v}^\prime,t)$.

For simplicity, the remainder of the paper will consider one dimensional neuronal models subject to external 
Poisson distributed spike trains,  each spike causing an instantaneous jump in the membrane potential (the case 
of more than one input, or step sizes $h$ that originate from a distribution of 
synaptic efficacies can be handled easily):
\begin{equation}
W(v^{\prime} \mid v) = \nu \delta (v^{\prime} - h - v),
\end{equation}
leading to:
\begin{equation}
  \frac{ d \rho}{dt} = \nu \left\{ \rho(v^\prime - v^\prime_h,t) -\rho(v^\prime,t) \right\} +r(t)\delta(v^\prime - v^\prime_{reset}),
  \label{eq-inhom}
\end{equation}
Here $r(t) = \int^{v^{\prime}_{th}}_{v^{\prime}_{th,h}} \rho(w,t) dw$; the $\delta$ peak reflects the reset of neurons that 
spiked to their reset potential.

Equation \ref{eq-inhom} is the Master equation of an inhomogeneous Poisson process. When considered over a 
sufficiently short time scale, the solution of this process can be approximated by a homogeneous one, for which 
efficient numerical and analytic methods are available \cite{dekamps2006}. The main technical problem associated 
with solving Eq. \ref{eq-inhom} is to find a grid that adequately represents the density, not only at $t=0$, 
but also at later times. Note that in $v^{\prime}$-space all evolution is due to synaptic input, as in the absence 
of noise $\frac{ d \rho(v^{\prime},t)}{dt} = 0$.

This idea works well for leaky-integrate-and-fire neurons \cite{dekamps2003}: 
\begin{equation}
  \tau \frac{dV}{dt} = -V  + I,
  \label{eq-lifchar}
\end{equation}
where $V$ is the membrane potential and $I$ an external (non-stochastic) current.
Here $v^{\prime}$-space slowly expands, which can be accommodated for by adding points to the grid representing 
the density profile during simulation. Occasionally one must rebin to curb the resulting growth of the grid 
(see \cite{dekamps2003} for details).

\begin{figure}[!ht]
\begin{center}
\includegraphics[width=0.9\textwidth]{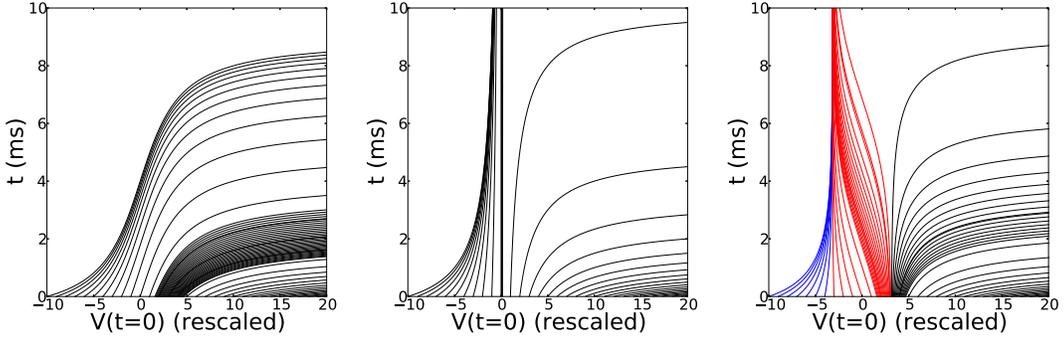}
\end{center}
\caption{\bf Characteristics for the values of $I$: $I = -10$  (top), $I = 0$  (middle) and $I = 10$  (bottom), on an interval $D = [-10, 10]$.}
\label{fig-charqif}
\end{figure}

In general neuronal dynamics is not slow compared to synaptic input, e.g. consider the 
quadratic-integrate-and-fire
model given by:
\begin{equation}
\tau \frac{dV}{dt} = V^{2} + I,
\label{eq-qif}
\end{equation}
where $V$ is a rescaled dimensionless variable. The characteristics can be calculated analytically 
(Table \ref{tab-qifchar}), and are shown in Fig. \ref{fig-charqif}. Their topology depends on the value of $I$. 
It is clear that some curves run away to infinity in finite time: these neurons are intrinsically  spiking.  
Upon reaching infinity, or some threshold potential $V_{th}$, these neurons are reintroduced in the system at 
$V=V_{reset}$.

\begin{table}[!ht]
  \begin{center}
  \begin{tabular}{||l|l|l||} \hline \hline
   $I > 0$ & $t = \frac{\tau}{\sqrt{I}}\left\{ \arctan \frac{V}{\sqrt{I}} - \arctan \frac{V_0}{\sqrt{I}} \right\}$ & $V(t) =  \sqrt{I} \tan \left\{ \sqrt{I} \frac{t}{\tau} + \arctan(\frac{V_0}{\sqrt{I}}) \right\}$ \\ \hline
  $I = 0$ & $t = \frac{\tau}{V_0} - \frac{1}{V}$ & $V(t) = \frac{V_0}{1-V_0 \frac{t}{\tau}}$ \\ \hline
  $I < 0$ & $t = \frac{\tau}{2 \sqrt{I}} \ln \left\{ (\frac{V- \sqrt{I}}{V+\sqrt{I}})(\frac{V_0 + \sqrt{I}}{V_0 -\sqrt{I}}) \right\}$ & $
V(t) = \sqrt{I} \frac{(\sqrt{I}+ V_0)e^{-2\sqrt{I} \frac{t}{\tau}} - (\sqrt{I}-V_0)}{(\sqrt{I}+V_0)e^{-2\sqrt{I} \frac{t}{\tau}} + (\sqrt{I}-V_0)} $ \\ \hline \hline
\end{tabular}
\end{center}
\caption{ \bf{QIF Characteristics}}

\begin{flushleft} The solutions to Eq. \ref{eq-dyn}, for different values of $I$.
\end{flushleft}
\label{tab-qifchar}
\end{table}

At first sight, it seems hard to find an adequate density representation without being forced to rapid and 
expensive rebinning  operations that must be applied in brief succession. However, by adopting a grid whose 
bin limits are defined by the characteristics rebinning operations can be avoided altogether. Fast dynamics 
will result in large bins, slow dynamics in small ones. Large bins, however, do not introduce inaccuracy, because
all density within the same bin will share a common fate. This is true even after reset. This observation is 
crucial and applies quite generally. For this reason we will explain it in some detail in the next section.
\footnote{The reset is a crucial difference between neuroscience and finance. It is one reason why analytic 
results for financial derivatives do not carry over to neuroscience and why analytic results there are almost 
unobtainable. The method must be able to handle reset, but will also work for
systems where individuals are not reintroduced.}

\subsection*{Representing fast dynamics}

\begin{figure}[!ht]
\begin{center}
\includegraphics[width=0.3\textwidth]{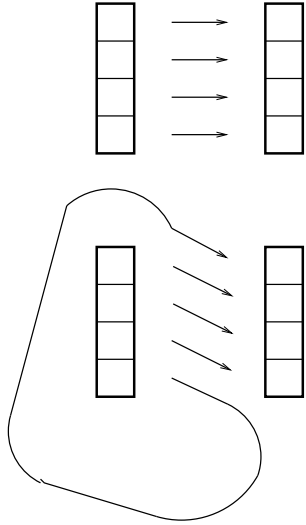}
\end{center}
\caption{ {\bf Modelling advection requires only a pointer update. } \emph{Top}: At time $t$ each probability bin
is associated with a density interval. \emph{Bottom}: Update at $t+1$.}
\label{fig-advection}
\end{figure}

Starting with a model of the form of Eq. \ref{eq-dyn}, we assume that $\frac{dV}{dt} > 0 $ everywhere 
on $(V_{min},V_{th})$, considering more general cases later on.

Consider now the initial value problem, posed by Equation \ref{eq-dyn} with boundary condition 
$V(t = 0) = V_{0}$. Assuming $V_0 = V_{min}$, the neuron will reach threshold $V_{th}$ in finite time $t_{period}$. 
The neuron will then be reintroduced at $V_{reset}$. For simplicity $V_{reset} = V_{min}$ will be assumed, for a 
practical implementation of the algorithm this is immaterial. Consider now a division of time 
interval $t_{period}$ into $N$ time steps:
\begin{equation}
t_{step} \equiv \frac{t_{period}}{N}
\end{equation} 

The density profile $\rho(v)$ for given time $t$ will be represented in a grid consisting of $N$ bins. We 
define the bin limits as follows: each bin $i$ ($i = 0, \cdots, N -1$) corresponds to a potential interval 
$[v_i, v_{i+1})$ in the following way: $v_0 = V_{min}$, and when $V(t = 0)= V_{min}$, $v_i$ is given by:
\begin{equation}
v_i \equiv V(i t_{step}), i = 0, \cdots, N,
\end{equation}
where $V(t)$ is the solution of Eq. \ref{eq-dyn}. Note that there are $N+1$ bin limits and that by 
definition $V_N \equiv V_{th}$. In general, the binning is non-equidistant. When a neuron spikes, for example, 
it traverses a considerable potential difference in a short period of time and the bins covering this traversal 
will be very large. The evolution of a population density profile defined on $(V_{min}, V_{th})$ can now be modeled 
as follows. A probability grid $\mathcal{P}$ is created of size $N$.  An array $\mathcal{V}$ contains the bin 
limits of $\mathcal{P}$. We define element $i$  of $\mathcal{V}$, $\mathcal{V}_i \equiv v_{i}$. Assume an initial 
density profile $\rho(v,0)$ is defined at $t=0$. We represent this profile by setting $P_{i}$ to:
\begin{equation}
  P_{i} \equiv \rho(v_{i},0)(v_{i+1} - v_{i}),
\end{equation}
so that $P_{i}$ approximates  the total probability between $V=v_{i}$ and $V=v_{i+1}$. Remarkably, having defined 
the contents of $\mathcal{V}$ and $\mathcal{P}$, they can remain constant, yet describe advection: the evolution 
of the density in the absence of synaptic input. To see this, consider the evolution over a period of time 
$t_{step}$. All neurons that previously had a potential between $V=v_i$ and $V=v_{i+1}$, will now have
a potential between $V=v_{i+1}$ and $V=v_{i+2}$, except for those that had a potential between 
$V=v_{N-1}$ and $V_{th}$ who now have a potential between $V=V_{min}$ and $V=v_1$. So, the relationship between 
$\mathcal{V}$ and $\mathcal{P}$ changes, but not the actual array contents
themselves. Specifically, the density profile at simulation time $t_{sim} = j t_{step}$ is given by:
\begin{equation}
  \rho(v_{i},t_{sim}) = \frac{P_{(i-j) \bmod N}}{v_{i+1} - v_{i}}        
\end{equation}
This is all that is required to represent the density profile in $v^{\prime}$-space; the computational overhead is 
negligible as an implementation of this idea only requires a pointer update without any need for shuffling data
around (see Fig. \ref{fig-advection}).

Equation \ref{eq-inhom} can be solved by assuming that during $t_{step}$ neurons only leave their bin due synaptic 
input.  It is also assumed that, although the magnitude of $h$ is dependent on $v^{\prime}$, it is constant during 
the small time $t_{step}$. The solution method is not really different from that described in 
\cite{sirovich2003,dekamps2003,dekamps2006}, but is complicated by the non-equidistance of the probability grid 
$\mathcal{P}$. Where in the original implementation only the magnitude of the synaptic efficacy had to be 
recalculated in $v^{\prime}$-space to find the transition matrix of the Poisson process, here a search is
needed to locate the density bin that will receive probability from a given density bin, as the
jump size $h^{\prime}$ becomes dependent on $v^{\prime}$ in $v^{\prime}$-space.
This issue is rather technical and is be explained in detail in 
{\bf ``Methods: Synaptic Input: Solving the Master Equation''}.

\begin{figure}[!ht]
\begin{center}
\includegraphics[width=\textwidth]{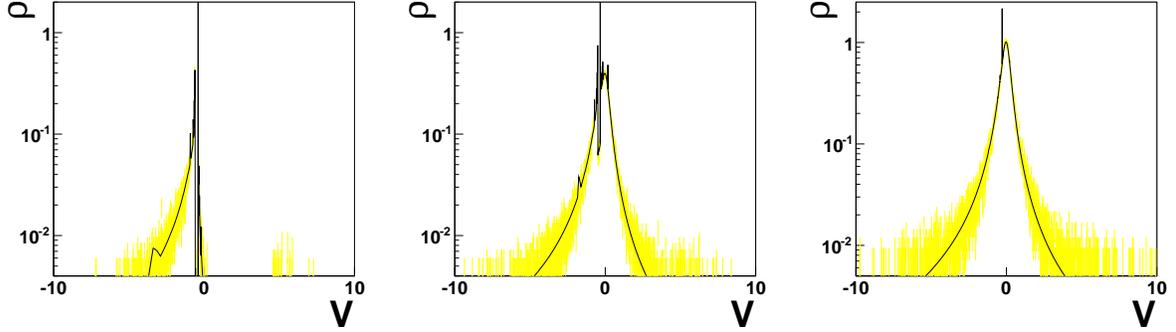}
\end{center}
\caption{{\bf Density profile of QIF population decorrelated by  
large synapses.} Line: population density; markers: Monte Carlo simulation.  }
\label{fig-large} 
\end{figure}

\subsection*{Switching Topologies: Current Compensation}
The idea is most easily illustrated for a Poisson process.
\label{sec-diffusion}
In one dimension, for a single synaptic input with postsynaptic efficacy $h$ receiving a single Poisson 
distributed input rate $\nu$, 
the population density equation is given
by \cite{omurtag2000}:
\begin{equation}
  \frac{\partial \rho}{\partial t} + \frac{\partial}{\partial v} (\frac{F(v) \rho(v)}{\tau}) = \nu \left\{ \rho(v-h) - \rho(v) \right\}
\end{equation}
Consider the limit $h \rightarrow 0$, such that $\nu h = \mbox{constant}$. A Taylor expansion of the right hand 
side up to order $h^2$ then gives:
\begin{equation}
\frac{\partial \rho}{\partial t} + \frac{\partial}{\partial v} \left\{  \frac{F(v) \rho}{\tau} + \rho h \nu - \frac{\nu h^2}{2} \frac{\partial \rho}{\partial v} \right\} = 0
\label{eq-protofp}
\end{equation} 
Now define:
\begin{align}
  \mu       & \equiv       \nu h \tau \nonumber \\
  \sigma^2  & \equiv  \nu h^2 \tau 
  \label{eq-musig}
\end{align}
and Eq. \ref{eq-protofp} becomes a Fokker-Planck equation:
\begin{equation}
\frac{\partial \rho}{\partial t} + \frac{\partial}{\partial v} \left\{ \rho  \frac{F(v)  + \mu}{\tau} - \frac{\sigma^2}{2 \tau} \frac{\partial \rho}{\partial v} \right\} = 0
\label{eq-fp}
\end{equation}
This equation describes the evolution of the density due to the deterministic dynamics, determined by $F(v)$ and 
an additive Gaussian white noise with
parameters $\mu$ and $\sigma$.

It is clear that Eq. \ref{eq-fp} is invariant under the transformation:
\begin{equation}
  \left\{ \begin{array}{rl} F(v)& \rightarrow F(v) + I_c \\ \mu & \rightarrow \mu - I_c \end{array} \right.
\end{equation}
for any constant current value $I_c$, i.e. one add a DC component to a neuron when a similar mean contribution
is subtracted from its input. This result is valid beyond the diffusion approximation, as it still holds
when higher than second order derivatives are considered in the procedure used to derive Eq. \ref{eq-fp}.

\begin{table}[!ht]
\caption{\bf{Default simulation parameters for QIF neurons.}}
\begin{center}
\begin{tabular}{||c|c||} \hline
  $N$ & 300 \\
  $\tau$   & $10^{-2}$ s\\
  $\tau_{ref}$ & 0 s \\
  $V_{min}$ & -10  \\
  $V_{max}$ &  10 \\
  $V_{reset}$ & -10 \\
 $I_c$     & 0.2 \\
  $\sigma_c$ & 0.01 \\
  $h_{diff}$ & 0.03 \\ \hline
\end{tabular}
\end{center}
\label{tab-sim}
\end{table}

\subsection*{Examples}
We will give a number of examples. For quadratic-integrate-and-fire neurons the  default simulation parameters are given Tab. 
\ref{tab-sim}.  In the remainder only deviations will be listed.

{\bf Example 1: quadratic-integrate-and-fire neurons with extremely large synapses} \\
\noindent Figure \ref{fig-large} demonstrates the validity of the method beyond the diffusion limit. 
A periodically firing group of neurons, firing in synchrony at the start of the simulation, is decorrelated by a 
low frequency (5 Hz) high impact ($h = 5$, this is half of the entire interval!) input. The
density profile is a slowly collapsing delta-peak, travelling along the characteristics of a periodically 
firing quadratic-integrate-and-fire neuron for several seconds, before reaching its steady state profile. 
We show the density profile in Fig. \ref{fig-large} at $t = 0.02, 0.12$ and $9.9$ s.

At no moment the population 
density is well described by a diffusion process, and the steady state firing rate deviates considerably from 
that predicted by numerical calculations based on the diffusion approximation. At $t = 0.12$ s, the density 
profile shows a fine structure that is indeed borne out by large-scale Monte Carlo simulations. We are not aware 
of  a method sensitive to the density gradient that could have modeled this accurately.

\begin{figure}[!ht]
\begin{center}
\includegraphics[width=\textwidth]{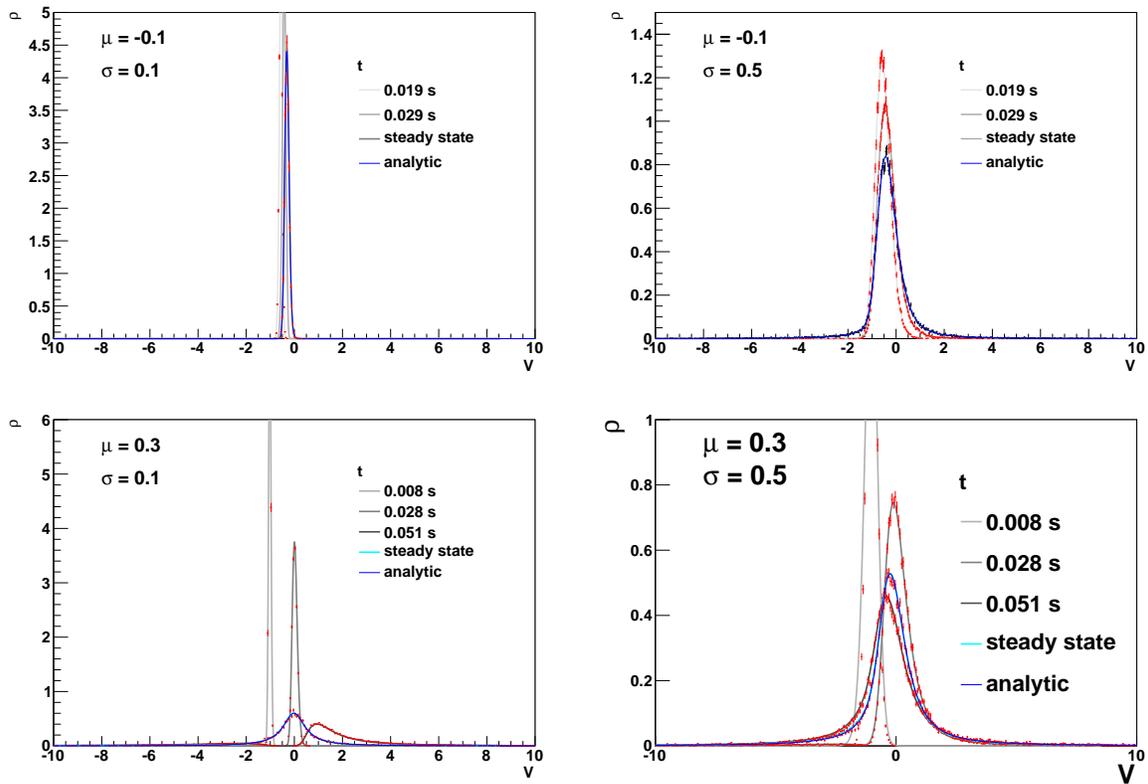}
\end{center}
\caption{{\bf Examples of the evolution of the density of a population of QIF neurons under the influence of Gaussian white noise input.}   }
\label{fig-density}
\end{figure}

{\bf Example 2: quadratic-integrate-and-fire neurons in the diffusion limit} \\
\noindent A single Poisson input, or a combination of one excitatory and one inhibitory Poisson input, can 
emulate a Gaussian white noise (see {\bf ``Methods: Emulation of Gaussian white noise''}). We will study both 
the density transient density profiles and the resulting 
firing rate response of the population (a neuron reaching threshold will emit a spike and reset to reset potential.
The population firing rate is fraction of neurons that in a time interval, divided by that time interval).

\begin{figure}
\begin{center}
  \includegraphics[width=\textwidth]{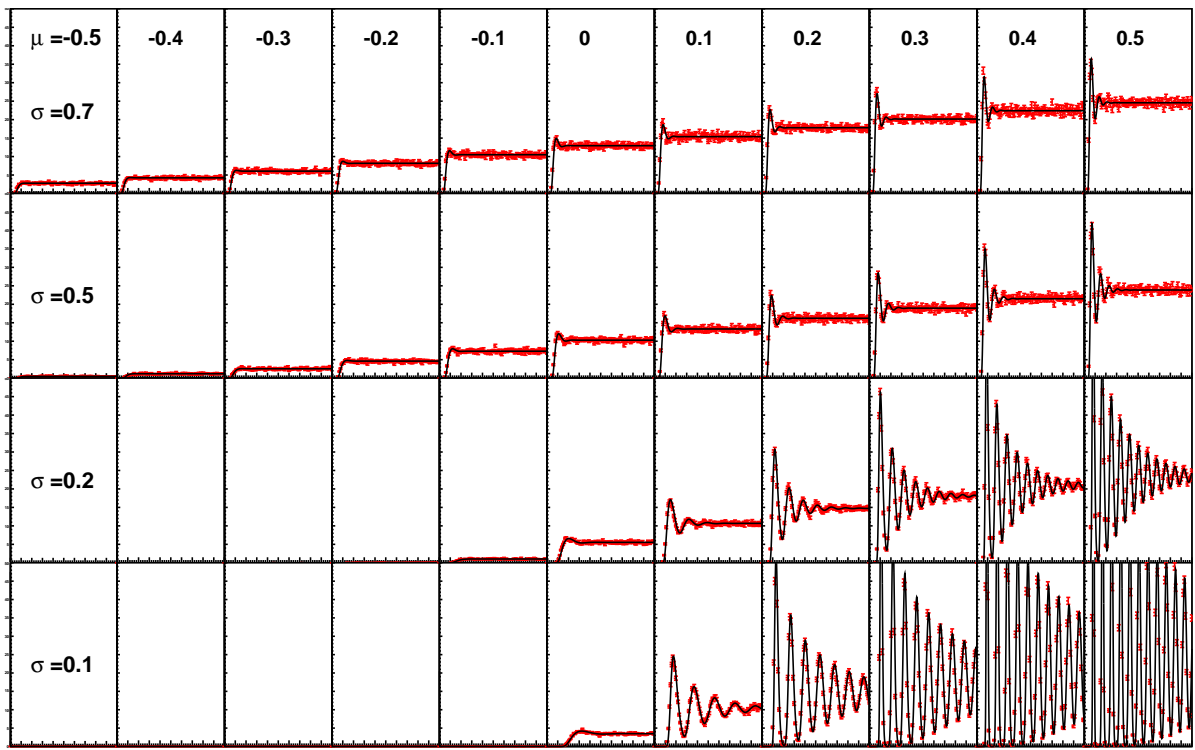}
\end{center}
\caption{{\bf  Transient firing rates inresponse to Gaussian white noise.  }  }
\label{fig-diffusionrates}
\end{figure}

In Fig. \ref{fig-density} the transient density profiles are shown for four different $(\mu, \sigma)$ 
combinations. The negative input, small noise distribution ($\mu = -0.1, \sigma = 0.1$) peaks at the stable 
equilibrium point. Almost no neurons are pushed across the second, instable
equilibrium point and this population does not fire in response to its input (see Fig. \ref{fig-diffusionrates}). 
For higher noise values, the peak is smeared and some neurons are pushed across the unstable, leading to an 
appreciable firing rate; a deterministic current with similar mean would evoke no response 
(see Fig. \ref{fig-diffusionrates}).

\begin{figure}
\begin{center}
  \includegraphics[width=\textwidth]{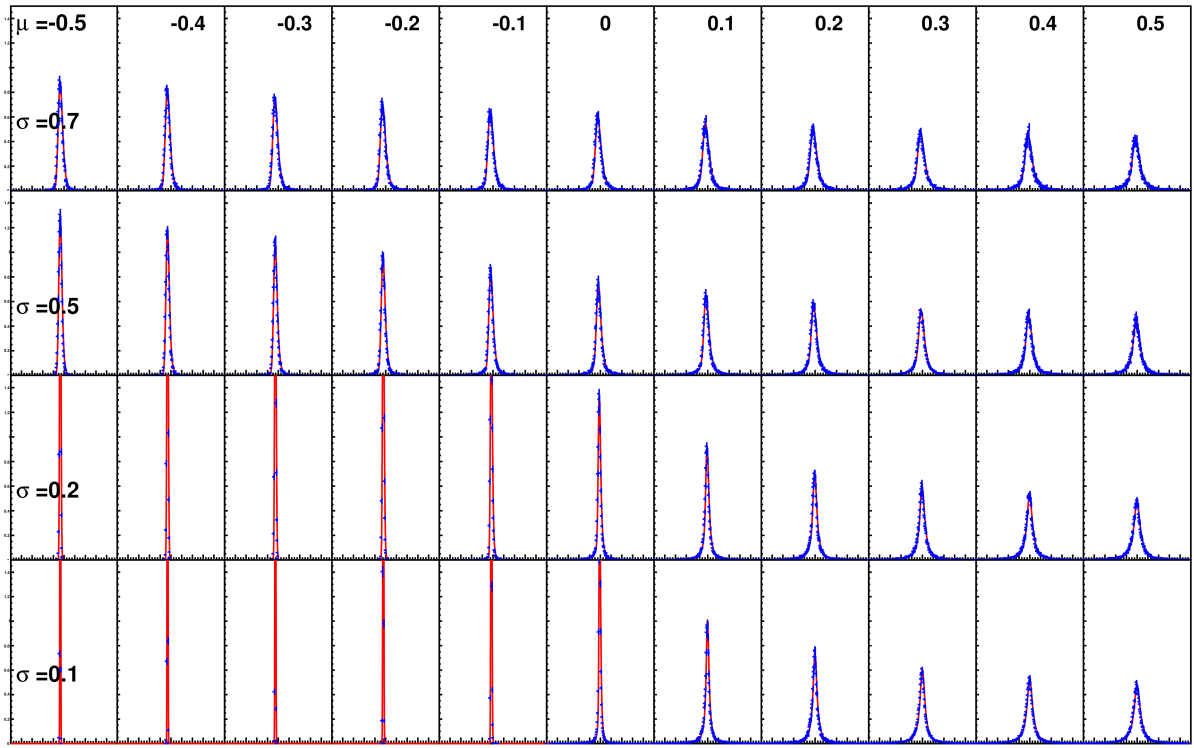}
\end{center}
\caption{{\bf  Steady 
state density profiles in response to Gaussian white noise.  } }
\label{fig-diffusionstates}
\end{figure}

A larger mean $\mu = 0.3$ evokes a clear response, and the peak of the density is clearly driven across 
$V = 0$. Unlike in Example 1, this peak will soon collapse. It will approach its steady state distribution but 
demonstrate small oscillations that are clearly visible in the firing rate
for a long time. At higher noise, the steady state is reached faster and the firing rates transients die out much 
faster (Fig. \ref{fig-diffusionrates}).

\begin{figure}
\begin{center}
  \includegraphics[width=0.5\textwidth]{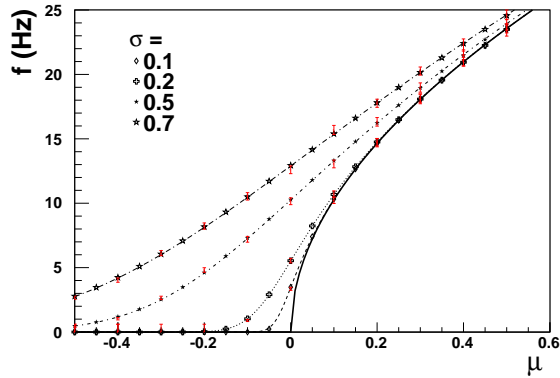}
\end{center}
\caption{{\bf Gain curves.} Lines: analytic steady state
firing rates. Red error bars: Monte Carlo simulation, black markers: population density results.}
\label{fig-gaindiffusion}
\end{figure}

Figures \ref{fig-diffusionstates}  and \ref{fig-diffusionrates} provide a summary of the diffusion results for $\mu \in [-0.5, 0.5]$ 
and $\sigma = 0.1, 0.2, 0.5, 0.7$. Figure \ref{fig-diffusionrates} shows the transient firing rates are shown and compared to Monte
Carlo simulations (red markers). Figure \ref{fig-diffusionstates}  shows the density profiles for 
large time $t = 9.9$ s, when profiles should assume their steady state (grey line, but barely visible
since Monte Carlo and analytic results agree and are overlaid). Markers show Monte Carlo results,
red lines are calculated directly from the diffusion approximation, using the integration
scheme from \cite{richardson2007}. Figure \ref{fig-gaindiffusion}  
shows the so-called gain curves, the steady state
firing rates that can be read of Fig. \ref{fig-diffusionrates}  in black markers. Monte Carlo results are indicated
by red markers, lines are calculated analytically from the diffusion approximation \cite{amit1997}. All
results are in excellent agreement with each other.

\begin{figure}[!ht]
\begin{center}
  \includegraphics[width=0.9\textwidth]{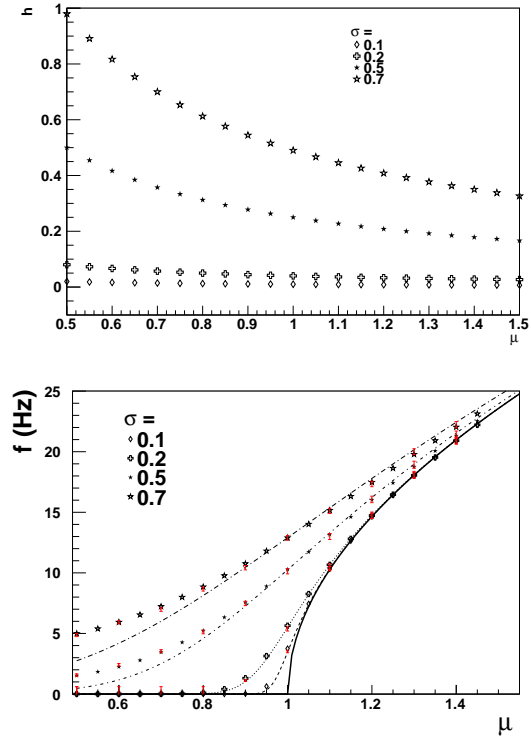}
\end{center}

\caption{{\bf For large synapses the gain curve changes.} \emph{Top}: Jump size for a single Poisson input 
emulating Gaussian white noise.  \emph{Bottom}: Monte Carlo (red markers), population
density results (black markers) and diffusion prediction (lines). }
\label{fig-deviation}
\end{figure}

{\bf Example 3: The gain curve is changed by large synapses.} \\
\noindent A  single Poisson input can only approximate Gaussian white noise as long as the variability
is much smaller than the mean of the signal. Figure \ref{fig-deviation} (top) shows the efficacy $h$ as a 
function of  $\mu$ for different $\sigma$
values, and hence indicate where deviations can be expected (namely for large $h$).  

Figure \ref{fig-deviation} shows a population receiving input from a  single Poisson input with firing rate and 
efficacy calculated using  Eq. \ref{eq-inverse}.  $\mu$ must not be
interpreted as the DC contribution of a Gaussian white noise input.

A value of $I = -1 (I_c = 1.1)$ was chosen and simulated both in Monte Carlo
and using the population density approach. Here, one expects the population to fire when the neurons' negative 
$I$ value is overcome, i.e. for $\mu > 1$.

The steady state values of the population density approach are in excellent agreement with the Monte Carlo,
but both deviate from the gain curves that are calculated in the diffusion approximation (Fig. \ref{fig-deviation}
(bottom)), precisely 
where this would have been expected according to Fig. \ref{fig-deviation} (top). 
Note, however, that for the first time 
it  was necessary to smear the synaptic efficacies slightly ($\sigma = 0.01$). In the absence of smearing, there 
would have been a  disagreement between Monte Carlo and data for low $\sigma$. This is due to the fact that the 
variability of the compensation current $I_c$ cannot be chosen arbitrarily low as this would entail artificially
high firing rates that would render the solution of the Master equation inefficient. A reasonable
minimum value is $\sigma = 0.01$. 

As can be seen from Fig. \ref{fig-deviation}, the jumps are really large for $\sigma = 0.7$ and low $\mu$. Indeed,
there the deviation from the diffusion approximation is substantial, but there is no visible disagreement between Monte 
Carlo and population density approach. This is all the more impressive because to model negative $I$ values, 
current compensation with $I_c = 1.1$ had to be used, and demonstrates the viability of the current
compensation approach.

When the steady state density profile obtained with the population density method is compared to that obtained 
from the Monte Carlo, again good agreement is found, while a small but significant deviation from the density 
profile calculated from the steady state, of the Fokker-Planck equation
Eq. \ref{eq-fp} can be seen for large $\sigma$ and low $\mu$ (Fig. \ref{fig-deviation} (top)).

{\bf Example 4: Universality: quadratic- and leaky-integrate-and-fire neurons are handled by the same algorithm.}\\
\noindent 
We use current compensation to turn leaky-integrate-and-fire neurons into spiking neurons. We add
a small constant current $I_c = 1.1$ as in Eq. \ref{eq-lifchar}. Without this current the characteristics
move away from threshold, with this current they cross threshold (compare Fig. \ref{fig-lifchar} left and right).

\begin{figure}[!ht]
\begin{center}
  \includegraphics[width=0.9\textwidth]{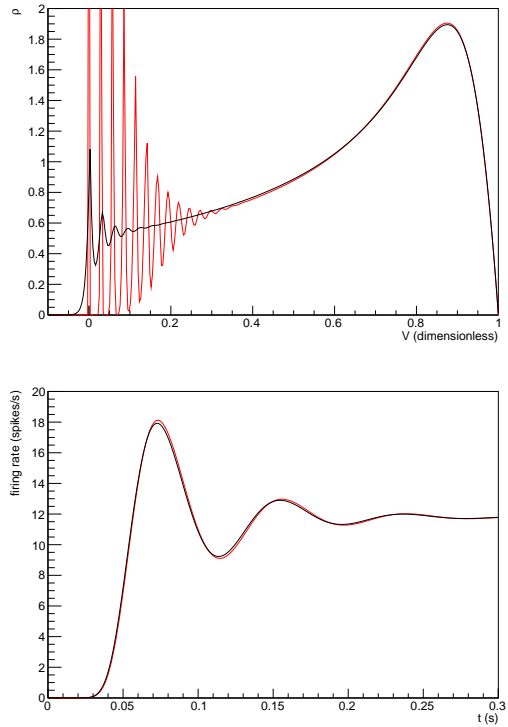}
\end{center}
\caption{ {\bf Algorithm works for both QIF and LIF neurons.}  \emph{Top}: 
steady state density profile for current compensation (black), 
dedicated LIF neuron implementation (red).
\emph{Bottom}: output firing rates.}
\label{fig-universal}
\end{figure}

A Gaussian white noise current emulated by Poisson input with mean $\mu = -1.1$ and small variability
($\sigma = 0.05$) is presented as input. This is combined with a single Poisson input of 800 Hz
with synaptic efficacy $h = 0.03$. This example was used in earlier studies as a benchmark 
\cite{omurtag2000,dekamps2003}.  In Fig. \ref{fig-universal} the steady state density profile and the firing
rate of the population are shown. They are compared to the simulation without current compensation, described
earlier \cite{dekamps2003}, using a density representation tailored to leaky-integrate-and-fire characteristics 
(red line).

The results are almost identical. In the original simulation the steep peaks are the consequence of a single
synaptic efficacy. Sirovich \cite{sirovich2003} has shown analytically that for the first peak the density
profile is discontinuous, for the second peak the first derivative is discontinuous, and so on. This is not
necessarily realistic in terms of neuroscience, but it demonstrates once more that the method can handle
the jagged density profiles that result from finite size jump processes.

In the current compensation version of the simulation, the small variability of the compensation current smears 
these peaks. The noise also slightly reduce the extremes of the firing rate (Fig. \ref{fig-universal}),
but otherwise the agreement is excellent. The effect of the smearing due to the compensation current is equivalent
to effects caused by spread in the synaptic efficacies \cite{amit1997}. Moreover, many brain systems
receive an unspecific background that must be modeled as noise anyway, so current compensation can be applied
quite generally in neuroscience.

Importantly, the only change that had to made to change over from quadratic- to leaky-integrate-and-fire neurons
is in the calculation of array $\mathcal{V}$, which is done during the initialization of the algorithm.

\begin{figure}[!ht]
\begin{center}
\includegraphics[width = 0.9\textwidth]{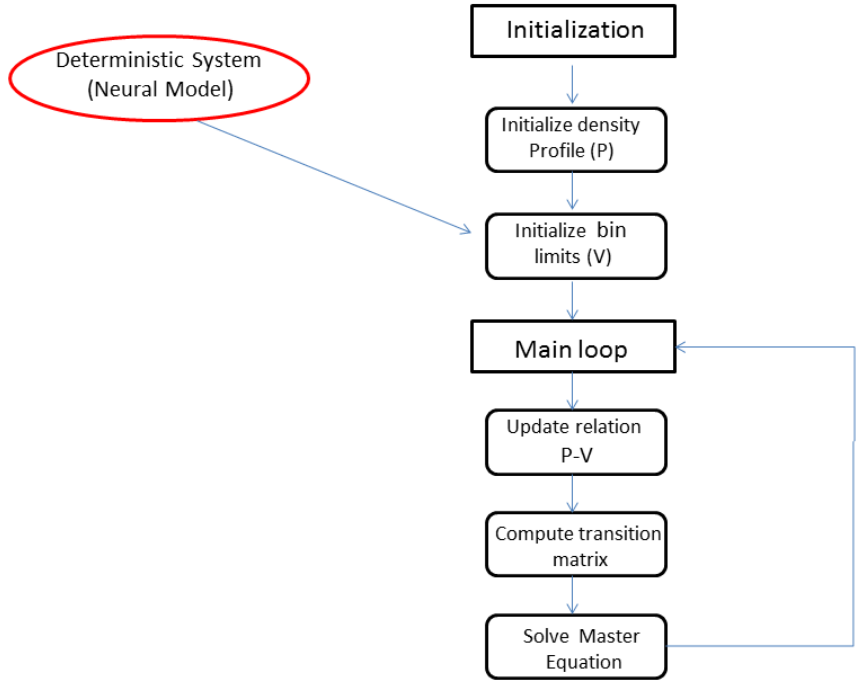}
\caption{{\bf High level overview of the approach.} Neural model (red)
only enters explicitly is in  $\mathcal{V}$.}
\label{fig-algo}
\end{center}
\end{figure}

\section*{Discussion}
We presented a method for numerically solving the differential Chapman-Kolmogorov equation for large jumps. 
We demonstrated
the validity of the approach by comparing population density results for neuronal populations receiving input 
from large synapses to Monte Carlo simulations, and neurons receiving input from small synapses to analytic or 
numerical results obtained in a diffusion approach. We presented two examples (Example 1 and 4) that demonstrate 
that the method can handle arbitrary jump sizes and the extremely jagged density profiles that result.

We also demonstrated that the algorithm is universal: it can handle \emph{any} one dimensional neural,
subject to the condition that a (pseudo) diffusive background of small variability is allowed. This
point is obvious mathematically, the transformation from the differential Chapman-Kolmogorov equation
to the Master equation of the noise process can always be performed \cite{dekamps2003}. The challenge is to find
a suitable density representation. 
Periodically spiking neurons yield a particularly
simple density representation, and using current compensation a non spiking neural model can be made to spike. In
practice the compensation current will always introduce some variability. Where variability is present
already, for example because the input is noise or because synaptic efficacies are variable, 
this extra variability can be absorbed and has no bearing on the simulation results.

The universality of the algorithm is reflected in the implementation. Although the neuronal model
defines the grid boundaries of the density representation - the content of grid $\mathcal{V}$, it does so before 
simulation starts. This means that neuronal models can be exchanged with great simplicity. To provide
a neural model is to provide the contents of grid $\mathcal{V}$. This is the only place were the neuronal
model enters the method explicitly. This important point is emphasized in Fig. \ref{fig-algo},
which provides a high level overview of the method.

We  demonstrated the method using a Poisson process and showed that if the jump size becomes
small, diffusion results are recovered. The use of Poisson statistics is appropriate in neural simulations.
An individual neuron will receive at most of the order of $10^5$ spikes per seconds. At such rates,
a Poisson Master equation can solved efficiently.  We found that 1 simulation second could be handled in between 
a 0.5 and 2.0 s real time, which is an order of magnitude faster than Monte Carlo - based on comparisons
to  leaky-integrate-and-fire \footnote{Quadratic-integrate-and-fire neurons were not implemented at the time
of writing. We performed Monte Carlo simulations by writing an event-driven simulator based on the analytic solutions of Table \ref{tab-qifchar}.} simulations in NEST \cite{nest} where a 10000 neuron population achieves real time ratios
of approximately 30. We found an excellent recovery 
of  diffusion results whenever the jump size is 5 \% of the interval for low noise and of the order of 10 \% for 
high noise conditions (Fig. \ref{fig-deviation}). 

It is conceivable that in other applications jump  sizes are smaller. Artificially small jump sizes would lead
to very high firing rates for a Poisson master equation. We find that adopting a jump size of less than 1 \% of 
the interval leads to inconveniently high firing rates. The method is clearly best used for finite jump sizes.

To study diffusion, it would be interesting to consider a finite difference scheme for solving Eq. \ref{eq-fp} 
with $F(v) = 0$. This would allow a direct comparison to diffusion results, while still benefitting from
the universality of this approach. For example, it would still be possible to compare diffusion in different 
neural models without the need to derive a separate Fokker-Planck equation for each case. Cox
and Ross \cite{cox1976} have shown that explicit finite difference schemes for solving equations of the type
\ref{eq-fp} correspond to a jump process with three possibilities: a move left, right or no move in the grid.
Brennan and Schwartz \cite{brennan1978} have shown that implicit finite difference scheme corresponds
to a jump process where jumps to every other position are allowed. For suitable parameters and 
in the limit of an infinitely fine grid both schemes converge to a diffusion process, where the implicit
scheme is unconditionally stable. These results are interesting because they suggest suitable Master equations 
to arrive at good diffusion approximations. In the context of neural systems this discussion is to some extent
academic as true diffusion conditions are not realized in the brain, but most analytic results on population
densities have been derived in the diffusion approximation, and a comparison is valuable.

The ideas presented here constitute an approach rather than an algorithm. There is an endless variation
of noise processes and dynamical systems that can be combined. Colored noise can be handled
by a two-step approach: the Master equation of a colored noise process can be simulated by 
leaky-integrate-and-fire population receiving white noise.

The technique was demonstrated on one dimensional neural populations. Such models are simple
enough to demonstrate the applicability of the technique to a wider audience. Whether it can be successfully
employed to more complex neural models is an open question. Fortunately, there is a tendency to move
from complex conductance-based models to simpler low dimensional effective model neurons, see e.g.
\cite{brette2005}. The population density approach still is competitive in two dimensions \cite{apfaltrer2006}.

We hope that the technique will find widespread application. Depending on the application area, the methods
described here may need adaptation. The current compensation technique may not work for all
applications as the extra variability introduced by the compensation current might not always be acceptable.
However, this is no fundamental objection to using this approach. Applying a combination of techniques
described here  and earlier \cite{dekamps2003,dekamps2006} useful density representations may still be
found, although their implementation may be limited to a specific topology of the deterministic flow.

\section*{Methods}

\subsection*{Solutions in absence of synaptic input}
Consider a neuron in the absence of synaptic input, but receiving a constant input 
current $I$. In general Equation \ref{eq-dyn} can be used to evolve the state of a neuron
that has potential $V_0$ at $t = 0$. For some cases analytic solutions exist. For leaky-integrate-and-fire neurons:
\begin{equation}
V(t) = I - (I-V_0)e^{-\frac{t}{\tau}}
\end{equation} 
For quadratic-integrate-and-fire neurons, the solutions of Eq. \ref{eq-qif} depend on the sign of $I$, and there are clear topological 
differences between solutions for different $I$. They
are given in Tab. \ref{tab-qifchar} and shown in Fig. \ref{fig-charqif}.

\subsection*{Derivation of the population density equation}
The population density approach can be motivated as follows and is not restricted to one dimensional neuronal 
models. To reflect that, this section
will consider the more general model:
\begin{equation}
\tau  \frac{d \vec{v}}{dt} = \vec{F}(\vec{v})
    \label{eq-multi}
\end{equation}
 Assume that $\mathcal{M}$ is an open subset of $\mathbb{R}^{n}$, where $n$ is the
dimensionality of the neuronal point model defined by Eq. \ref{eq-multi}. 
Assume that the neuronal state is given by  point $\vec{v}$ in $\mathcal{M}$. Further, assume
that a density function $\rho(\vec{v},t)$ is defined on $\mathcal{M}$ and that $\rho(\vec{v},t)d \vec{v}$ 
represents the probability for a given neuron to be in the state space element $d \vec{v}$ at time $t$. Consider 
the evolution of neuronal states for neurons
in $d \vec{v}$ during a short period of time $dt$. If external input is ignored, then under the influence of neuronal dynamics all neurons, and no others,
that were in $d \vec{v}$ at time $t$ are in volume element $d \vec{v}^{\prime}$ at time $t^{\prime}$,
with $\vec{v}^{\prime}$ and $t^{\prime}$ defined by:
\begin{equation}
  \left\{ 
  \begin{array}{rl}
    t \rightarrow & t^{\prime} = t + dt \\
    \vec{v} \rightarrow &\vec{v}^{\prime} = \vec{v} + \vec{dv} = \vec{v} + \frac{\vec{F}}{\tau}dt 
  \end{array}
\right.
\end{equation}
Conservation of probability requires:
\begin{equation}
\rho( \vec{v}, t)d \vec{v} = \rho( \vec{v}^{\prime}, t^\prime) d \vec{v}^{\prime}.
\label{eq-conserv}
\end{equation}
Using:
\begin{equation}
\rho(\vec{v}^{\prime},t^{\prime}) = \rho(\vec{v},t) + dt \frac{\partial \rho(\vec{v},t)}{\partial t} + d \vec{v} \cdot \frac{\partial \rho(\vec{v},t)}{\partial \vec{v}}, 
\end{equation}
Equation \ref{eq-conserv} becomes:
\begin{equation}
\left\{ \rho(\vec{v},t) + d \vec{v}\cdot \frac{\partial \rho(\vec{v}, t)}{\partial \vec{v}}  + dt \frac{ \partial \rho( \vec{v}, t)}{\partial t} \right\} d \vec{v}^{\prime} = \rho( \vec{v}, t) d \vec{v}
\end{equation}
Since
\begin{equation}
d \vec{v}^{\prime} = \mid J \mid d \vec{\vec{v}},
\end{equation}
with:
\begin{equation}
\mid J \mid = \mid  \frac{\partial \vec{v}^{\prime}}{\partial \vec{v}} \mid = \mid \mathbb{I} + \frac{1}{\tau}\frac{\partial \vec{F}( \vec{v})}{\partial \vec{v}}dt \mid + O(dt^2),
\end{equation}
so that, collecting all terms up to $O(dt)$:
\begin{equation}
\left\{\frac{1}{\tau} \frac{\partial \vec{F}( \vec{v} )}{\partial \vec{v}}\rho +\frac{\partial \rho} {\partial t} + \frac{\vec{F}(\vec{v})}{\tau} \cdot\frac{\partial \rho}{\partial \vec{v}} \right\} d \vec{v} = 0.
\end{equation}
Since this equation must hold for infinitesimal but arbitrary volumes $d \vec{v}$, it follows that:
\begin{equation}
 \frac{\partial \rho} {\partial t} + \frac{1}{\tau}\frac{ \partial }{\partial \vec{v}} \cdot (\vec{F}\rho)  = 0
\label{eq-advection}
\end{equation}
Equation \ref{eq-advection} is an advection equation for probability flow under the influence of neuronal dynamics.
When noise is also considered the equation becomes:
\begin{equation}
  \frac{ \partial \rho }{ \partial t } + \frac{1}{\tau}\frac{\partial}{ \partial \vec{v}} \cdot ( \vec{F} \rho ) = \int_{M} d \vec{w} \left\{ W(\vec{v} \mid \vec{w})\rho(\vec{w}) - W(\vec{w} \mid \vec{v}  )\rho(\vec{v}) \right\}. 
\label{eq-convection}
\end{equation}
This is an example of what Gardiner calls the differential Chapman-Kolmogorov equation \cite{gardiner1997}.
In the remainder only Poisson processes will be considered, but the method is applicable to any noise process that can be modeled by a Master equation.
For a single Poisson input $W(\vec{v}^{\prime} \mid \vec{v})$ is given by:
\begin{equation}
  W(\vec{v^{\prime}} \mid \vec{v} ) = \nu \delta (v^{\prime}_0 - v_0 - h) 
\end{equation}
Here $\nu$ is the rate of the Poisson process, it is assumed that one of the neuronal state variables, $v_0$ instantaneously transitions
from $v_0$ to $v_0 + h$, where $h$ is the synaptic efficacy. In a one dimensional neuronal model $v_0$ is $V$, the membrane potential.
Clearly, not all synapses in a population have a single value, and often one assumes:
\begin{equation}
  W(\vec{v^{\prime}} \mid \vec{v} ) = \nu \int dh p(h)\delta (v^{\prime}_0 - v_0 - h), 
  \label{eq-transit}
\end{equation}
where $p(h)$ is often assumed to be Gaussian \cite{omurtag2000,amit1997}. In a network where many populations connect to any given population,
multiple contributions of the form Eq. \ref{eq-transit} must be included - one for each separate input. In principle, each input must be modeled separately,
but often many inputs can be subsumed into a single one using the central limit theorem, leading to a Gaussian white noise input contribution. 
Sometimes kernels $p(h)$ of separate inputs overlap and can be integrated into one input contribution.

\begin{figure}[!ht]
\begin{center}
\includegraphics[width=0.5\textwidth]{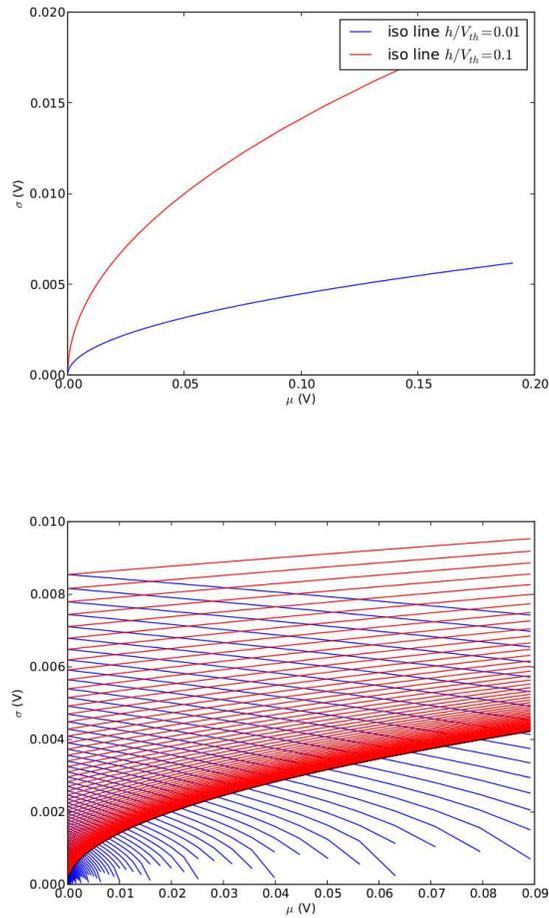}
\end{center}
\caption{{\bf Emulation of a Gaussian white noise with Poisson inputs.} Left: isoline 
of $h = 0.01 V_{th}, h= 0.1 V_{th}$.  Right:  synaptic 
efficacies are fixed at 3 \% of $V_{th}$. Blue  (red) lines are isolines of frequency of   excitatory 
(inhibitory) inputs.}
\label{fig-fpemulation}
\end{figure}

\begin{figure}[!ht]
\begin{center}
\includegraphics[width=0.6\textwidth]{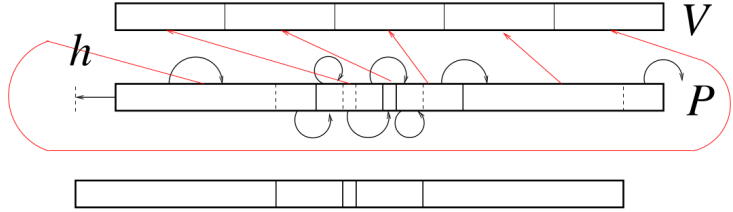}
\end{center}
\caption{ {\bf The Master equation follows from translating the contents of grid $\mathcal{V}$ the synaptic efficacy.}  }
\label{fig-master}
\end{figure}

\subsection*{Emulation of Gaussian white noise and the breakdown of the diffusion approximation}
\label{sec-breakdown}
It is evident that by choosing a small efficacy $h$ and large input rate $\nu$, the Poisson input 
approximates additive Gaussian white noise. These parameters
must be chosen such that they provide the correct $\mu$ and $\sigma$. Inverting Eq. \ref{eq-musig} gives:
\begin{align}
  h & = \frac{\sigma^2}{\mu} \nonumber\\
  \nu & = \frac{\mu^2}{ \tau \sigma^2}
  \label{eq-inverse}
\end{align}
A single Poisson input can only emulate a Gaussian white noise when $h$ is small compared to the relevant 
potential difference, e.g. for leaky-integrate-and-fire neurons when $\mid h \mid  \ll V_{th}$, or $\mid h \mid \ll I$ for quadratic-integrate-and-fire neurons, 
i.e. not for all values of ($\mu$,$\sigma$) can values for $h$ and $\nu$ be found so that the diffusion regime is 
reliably reproduced. This is borne out by Fig. \ref{fig-fpemulation} (left). Here a neuron is considered with 
$\tau = 10$ ms, $V_{th} (\equiv \theta) = 20$ mV. For
a $(\mu, \sigma)$ plane isolines of $h$ are plotted: one where $h$ is precisely at 1\% of the 
threshold potential, one where it is at 10\%.

It is clear that a single Poisson input can emulate a broad range of $\mu$ values, 
but only at low $\sigma$. Using two inputs, one excitatory and one inhibitory, one can fix the synaptic 
efficacies at a low value compared to $V_{th}$. If this value is $J$ one finds:
\begin{align}
  \mu &    = \tau J \nu (\nu_e - \nu_i) \nonumber \\
  \sigma^2 &  = \tau \nu J^2(\nu_e + \nu_i)
\end{align}

One then has two frequencies $\nu_e$ and $\nu_i$
as free parameters. This setup covers most of the $(\mu, \sigma)$ plane, except for values at low $\sigma$, 
where the input frequency of the inhibitory input falls below 0. In Fig. \ref{fig-fpemulation} only frequency 
isolines for positive input frequencies are shown. The boundary where the inhibitory input rate falls below 0 is 
forbidden for two inputs, but this is precisely the area covered by a single input (clearly if $\nu_i = 0$,
we have a single excitatory input). The arguments presented here cover a positive range of $\mu$, similar 
arguments can be made for a net inhibitory input.

In summary, one needs either one or two Poisson inputs to emulate a Gaussian white noise. In particular, 
this implies that we can use a single excitatory input as a model for the gradual breakdown of the diffusion 
limit: if we increase $\sigma$ and calculate the synaptic efficacy using Eq. \ref{eq-inverse},
we should see deviations from the diffusion limit as $h$ can no longer be considered to be small.

This is shown in Fig. \ref{fig-deviation} which shows the step size for a single Poisson input given 
$\mu$, $\sigma$. For $\sigma$ comparable to $\mu$ this step size is not necessarily small compared to the 
potential scale on which the neuronal model is defined and deviations of the
diffusion approximation can be expected.

Whenever a Gaussian white noise with very small variability is required, this 
must be delivered by a single Poisson input. The firing rate is then inversely proportional to the jump
size. For very low variability a numerical solution of the Poisson process may become inefficient.

\subsection*{Synaptic input: solving the master equation}
Now consider a single Poisson input. The evolution of the density profile is straightforward in principle 
but complicated by the bins being non-equidistant. Formally the noise process can
be represented as a transition matrix. Figure \ref{fig-master} how this matrix can be determined:
the contents of grid $\mathcal{V}$ are shifted by an amount $h$, where $h$ is the synaptic efficacy.
Probability moves from potential $V-h$ to $V$ as consequence of an input spike. In large bins this means
that neurons stay mostly within the same bin. Neurons that are in small bins may end up several bins higher.
Below we will examine this process in detail.

\begin{figure}[!ht]
\begin{center}
\includegraphics[width=\textwidth]{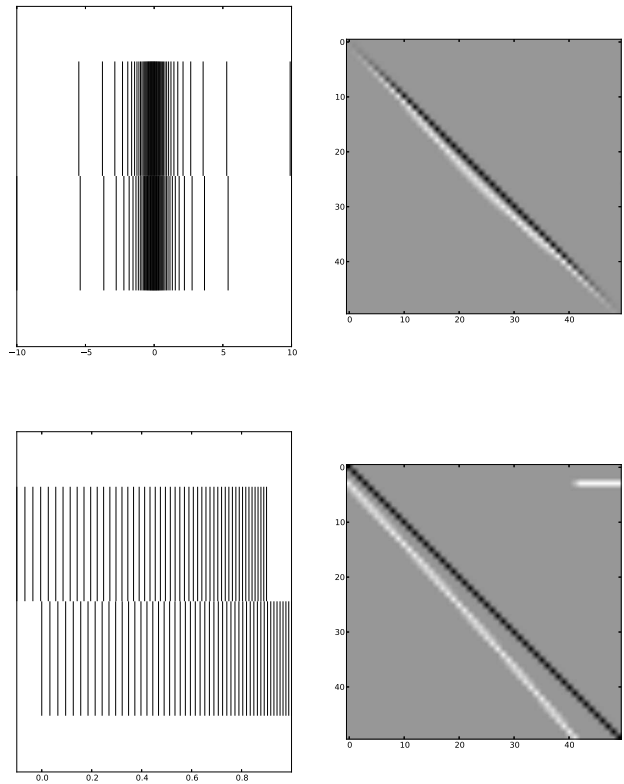}
\end{center}
\caption{ {\bf Transition matrices for QIF (top)  and LIF neurons (bottom).} 
Loss of probability is indicated by black, gain by white.  } 
\label{fig-transfer}
\end{figure}

To determine the Poisson Master equation for excitatory input $(h > 0)$, consider that every neuron arriving in the interval $D_i = [v_i, v_{i+1})$ must have come from
$D_{i,h} =[ \max( v_i - h, V_{min}), v_{i+1}-h)$. 
Consider, the $N+1$ integers  defined by:
\begin{equation}
  p_i \equiv
\left\{
   \begin{tabular}{ll} 
     $v_i -h < v_0$:& -1 \\
     $v_i -h \ge v_0$:& $\underset{k}{\argmin}\ v_{k}  \le  v_i-h    < v_{k+1}$  
     \end{tabular}
\right.
\end{equation}
Now consider $i = 0 \cdots N-1$; if $p_i = p_{i+1}$, $\alpha_{i,p_i}$ is given by:
\begin{equation}
\alpha_{i,p_i} \equiv
 \frac{v_{i+1} - v_i}{v_{p_i +1} - v_{p_i}};
\end{equation}
For $p_i = -1$, $\alpha_{i,l} = 0$.
If  $p_i < p_{i+1}$, $\alpha_{i,l}$ is given by:
\begin{equation}
  \alpha_{il} \equiv \left\{
  \begin{tabular}{rl}
   $l= p_i$:          & $\theta_{l} \frac{v_{p_i+1} - (v_i - h)}{v_{p_i+1}-v_{p_i}}$ \\
  $p_i < l < p_{i+1}$: & 1 \\
  $l = p_{i+1}$:       & $\frac{v_{i+1}-h - v_{p_{i+1}}}{v_{p_{i+1}+1} - v_{p_{i+1}}}$
  \end{tabular}
 \right.
\end{equation}
with:
\begin{equation} 
\theta_l \equiv \left\{
   \begin{tabular}{rl}
   $l <0$: & 0 \\
   $l \ge 0$: & 1 
   \end{tabular}
\right.
\end{equation}
 Finally,
\begin{equation}
  \beta_i \equiv \left\{
  \begin{tabular}{rl}
    $i < p_N$: & 0 \\
    $i = p_N$: & $\frac{v_{p_N + 1} - v_N + h}{v_{p_N +1} - v_{p_N}}$ \\
    $i > p_N$: & 1.
    \end{tabular}
\right.
\end{equation}
From this the Master equation at $t=kt_{step}$ follows:
\begin{align}
\frac{ d P_{i-k \bmod N}}{dt}  = & -P_{i-k \bmod N} + \sum^{p_{i+1}}_{l=p_i}\alpha_{i,l}P_{l-k  \bmod N} \\
                            + & \delta_{i R}(\sum^{N-1}_{l=0} \beta_l P_{l-k \bmod N}),
\label{eq-mastere}
\end{align}
where $R$ is defined by:
\begin{equation}
  R \equiv 
  \underset{k}{\argmin}\ v_{k}  \le  V_{reset}      \le v_{k+1}
\end{equation}
and $\delta$ is the Kronecker $\delta$. The term including it reflects the re-entry of neurons at the reset potential once they have spiked.

For inhibitory input $(h <0)$, the formulae are almost identical as now $v-h$ refers to a higher potential. Again, define
\begin{equation}
  \tilde{p}_i \equiv
\left\{
   \begin{tabular}{ll} 
     $v_i -h > v_N$:& $N$ \\
     $v_i -h \le v_N$:& $\underset{k}{\argmin}\ v_{k}  \le  v_i-h    < v_{k+1}$  
     \end{tabular}
\right.
\end{equation}
Again consider $i = 0 \cdots N-1$; if $\tilde{p_i} = \tilde{p}_{i+1}$, $\tilde{\alpha}_{i,\tilde{p}_i}$ is given by:
\begin{equation}
\tilde{\alpha}_{i,p_i} \equiv
 \frac{v_{i+1} - v_i}{v_{\tilde{p}_i +1} - v_{\tilde{p}_i}};
\end{equation}
if  $p_i < p_{i+1}$, $\tilde{\alpha}_{i,l}$ is given by:
\begin{equation}
  \tilde{\alpha}_{il} \equiv \left\{
  \begin{tabular}{rl}
   $l= \tilde{p}_i$:          & $\frac{v_{\tilde{p}_i+1} - (v_i + h)}{v_{\tilde{p}_i+1}-v_{\tilde{p}_i}}$ \\
  $\tilde{p}_i < l < \tilde{p}_{i+1}$: & 1 \\
  $l = \tilde{p}_{i+1}$:       & $\theta_{N-1-\tilde{p}_i + 1}\frac{v_{i+1}+h - v_{\tilde{p}_{i+1}}}{v_{\tilde{p}_{i+1}+1} - v_{\tilde{p}_{i+1}}}$
  \end{tabular}
 \right.
\end{equation}
Finally, $\gamma_i$ is defined as follows:
\begin{equation}
\gamma_i \equiv \left\{
\begin{tabular}{rl}
  $i < p_0$: & 1 \\
  $i = p_0$: & $\frac{ v_{p_0 + 1} - (v_0 + h) }{v_{p_0 + 1} - v_{p_0}}$ \\
  $i > p_0$: & 0
\end{tabular}
\right.
\end{equation}
The Master equation for an inhibitory input is:
\begin{equation}
  \frac{d P_{i - k \bmod N}}{dt} = -(1 - \gamma_i) P_{i-k \bmod N} + \sum^{\tilde{p}_{i+1}}_{l=\tilde{p}_i} \tilde{\alpha}_{i,l}P_{l-k \bmod N}
  \label{eq-masteri}
\end{equation}

Note that Eqs. \ref{eq-mastere} and \ref{eq-masteri} are of the form:
\begin{equation}
\frac{d \vec{P} }{dt } = \mathcal{M}^{(k)} \vec{P},
\label{eq-mastertot}
\end{equation}
where $\mathcal{M}^{(e)}$ is the transition matrix resulting from adding individual input contributions of the form \ref{eq-mastere} and \ref{eq-masteri}. 
\subsection*{Solving the master equation}
Equation \ref{eq-mastere} and \ref{eq-masteri} were solved numerically using Runge-Kutta-Fehlberg integration. This is not optimal, but is very
straightforward to implement, and yields a performance that is faster than Monte Carlo simulation by at least one order of magnitude. Note
that in Eq. \ref{eq-mastere} one simply ignores the zero elements of $\mathcal{M}^{(e)}$. The calculation of the time derivatives then scales as
$O(N)O(N_{input})$, where $N_{input}$ is the number of inputs that need to considered. In many neural systems, such as cortex it can be assumed
that most inputs deliver a background that can be treated as a single noisy input \cite{apfaltrer2006,amit1997}

\subsection*{Different neuronal models: different transition matrices}
Although the noise process and the neuronal dynamics are independent, the neuronal dynamics determines the 
representation of the density profile. The bin boundaries of the density profile in $v$-space, $\mathcal{V}$ are 
given by the evolution of the neuronal state, and therefore the transition matrix is directly dependent on the 
neuronal model.
This is illustrated in Fig. \ref{fig-transfer}, which shows the transition matrix for identical $h, \nu$ for quadratic-integrate-and-fire 
and leaky-integrate-and-fire neurons. First, it should be observed that to accommodate for spiking behaviour, the state space of the 
quadratic-integrate-and-fire neuron is much larger, for the leaky-integrate-and-fire neuron it is limited by the threshold, $V_{th}$.
Second, the quadratic-integrate-and-fire neuron is a truly spiking model: once it spikes, its state runs away to infinity in finite time. 
The leaky-integrate-and-fire neuron runs over threshold towards the externally applied current in infinite time.

It is clear that probability lost from a bin will end up in a higher bin, i.e. a bin corresponding to a higher 
potential, as expected for an excitatory event. For leaky-integrate-and-fire neurons the same is done. The differences between a 
spiking neuron model and a non-spiking model are clearly visible. The outer bins  of the quadratic-integrate-and-fire grid are huge, 
because the potential difference bridged during a simulation step $t_{step}$ is large during a spike. Relative to 
their size the outer bins are barely shifted with respect to each other, meaning that most probability originating
 from this bin during a synaptic event will end up there again.
This makes intuitive sense: during a spike the influence of synaptic input is negligible. Indeed the transition 
matrix shows that no probability transfer takes place in the outer bins for quadratic-integrate-and-fire neurons. For leaky-integrate-and-fire neurons, 
the picture is different: neurons will mainly pass threshold due to synaptic input. After
crossing threshold, their potential will be reset. The influence of the reset is clearly visible in the 
leaky-integrate-and-fire transition matrix as a horizontal line
in the row corresponding to the reset bin.

\section*{Acknowledgments}


\newpage



\end{document}